\documentclass[aps,twocolumn,groupedaddress,showpacs]{revtex4}

\begin{document}
\bibliographystyle{apsrev}


\title{Artificial light and quantum order in systems of screened dipoles}


\author{Xiao-Gang Wen}
\homepage{http://dao.mit.edu/~wen}
\affiliation{Department of Physics, Massachusetts Institute of Technology,
Cambridge, Massachusetts 02139}


\date{Sept. 2002}

\begin{abstract}
The origin of light is a unsolved mystery in nature. Recently, it was
suggested that light may originate from a new kind of order - quantum
order. To test this idea in experiments, we study systems of screened
magnetic/electric dipoles in 2D and 3D lattices.  We show that our models
contain an artificial light -- a photon-like collective excitation.  We
discuss how to design realistic devices that realize our models. We show that
the ``speed of light'' and the ``fine structure constant'' of the artificial
light can be tuned in our models. The properties of artificial atoms (bound
states of pairs of artificial charges) are also discussed. The existence of
artificial light (as well as artificial electron) in condensed matter systems
suggests that elementary particles, such as light and electron, may not be
elementary.  They may be collective excitations of quantum order in our
vacuum.  Our models further suggest that a gauge theory is a string-net
theory in disguise.  Light is a fluctuation of string-nets and
charges are the ends of open strings (or nodes of string-nets).  
\end{abstract}

\maketitle


\section{ Introduction}

What is light? Where light comes from? Why light exists?  Every one probably
agrees that those are fundamental questions. But one may wonder if they are
scientific questions, philosophical questions, or even religious question?
Before answering those questions and the questions about the questions, we
would like to ask three more questions: What is phonon? Where phonon comes
from?  Why phonon exists? \footnote{Note that the term phonon always refers to
gapless phonon in solid in this paper.} We know that those are scientific
questions and we know their answers. Phonon is a vibration of a crystal.
Phonon comes from a spontaneous translation symmetry breaking. Phonon exists
because the translation-symmetry-breaking phase actually exists in nature.

It is quite interesting to see that our understanding of a gapless excitation
- phonon - is rooted in our understanding of phases of matter.  According to
  Landau's theory,\cite{L3726} phases of matter are different because they
have different broken symmetries.  The symmetry description of phases is very
powerful. It allows us to classify all possible crystals. It also provide the
origin for gapless phonons and many other gapless
excitations.\cite{N6080,G6154}

However, light, as a $U(1)$ gauge boson, cannot be a Nambu-Goldstone mode
from a broken symmetry.  Therefore, unlike phonon, light cannot originate from
a symmetry breaking state.  This may be the reason why we treat light
differently than phonon. We regard light as an elementary particle and phonon
as a collective mode.  

However, if we believe in the equality between phonon and light
and if we believe that light is also a collective mode of a particular
``order'' in our vacuum, then the very existence of light implies a new kind
of order in our vacuum.  Thus, to understand the origin of light,
we need to deepen and expand our understanding of phases of matter. We need to
discover a new kind of order that can produce and protect light.

After the discovery of fractional quantum Hall (FQH)
effect,\cite{TSG8259,L8395} it became clear that the Landau's symmetry
breaking theory cannot describe different FQH states, since those states all
have the same symmetry. It was proposed that FQH states contain a new kind of
order - topological order.\cite{Wtoprev} The concept of topological order was
recently generalized to quantum order\cite{Wqoslpub,Wqogen} that is used to
describe new kind of orders in gapless quantum states. In particular, we used
quantum order and its projective symmetry group (PSG) description to classify
over one hundred different spin liquids that have the same
symmetry.\cite{Wqoslpub} Intuitively, we can view quantum/topological order as
a description of pattern of quantum entanglements in a quantum
state.\cite{Wqogen} The pattern of quantum entanglements, being described by
complex wave function, is much richer than pattern of classical
configurations. 

We know that the fluctuations of pattern of classical configurations (such as
lattices) lead to low energy collective excitations (such as phonons).
Similarly, the fluctuations of pattern of quantum entanglement also lead to
low energy collective excitations. However, collective excitations from
quantum entanglement can be gapless gauge
bosons\cite{DDL7863,W7985,BA8880,AM8874,RS9173,Wsrvb,WWZcsp,Wlight,MS0204}
and/or gapless fermions.  The fermions can even appear from pure bosonic
models on lattice.\cite{RK8876,RS9173,Wsrvb,K9721,Wqoslpub,Wlight,Wqoexct}

If we believe in quantum order, then the three questions about light will be
scientific questions. Their answer will be (A) light is a fluctuation of
quantum entanglement, (B) light comes from the quantum order in our vacuum and
(C) light exists because our vacuum contains a particular entanglement (\ie a
quantum order) that supports $U(1)$ gauge fluctuations.

According to the picture of quantum order, elementary particles (such as
photon and electron) may not be elementary after all. They may be collective
excitations of a bosonic system. Without experiments at Planck scale, it is
hard to prove or disprove if photon and electron are elementary particles or
not. However, one can make a point by showing that photon and electron can
exist as collective excitations in certain lattice bosonic models. 
So  photon and electron do not \emph{have} to be elementary particles.

The emerging gauge fluctuations (also called dynamically generated gauge
fields) has a long history. Dynamically generated $U(1)$ gauge field has been
introduced in quantum disordered phase of 1+1D $CP^N$
model.\cite{DDL7863,W7985} The $U(1)$ gauge field have also been found in the
slave-boson approach to spin liquid states of $SU(2)$ and $SU(N)$ spin
models on 2D square lattice.\cite{BA8880,AM8874} The slave-boson approach not
only introduces a $U(1)$ gauge field, it also introduces gapless fermion
fields.  However, due to the confinement of the $U(1)$ gauge field in 1+1D and
1+2D, non of the above gauge field and gapless fermion fields lead to gapless
gauge bosons and gapless fermions that appear as low energy physical
quasiparticles.  This led to an opinion that the $U(1)$ gauge field and the
gapless fermion fields are not real and are just a unphysical artifact of the
``unreliable'' slave-boson approach.  Thus the key to find dynamically
generated gauge boson is not to write down a Lagrangian that contain
\emph{gauge fields}, but to show that \emph{gauge bosons} actually appear in
the physical low energy spectrum.  Here, we would like to stress that ``gauge
fields'' and ``gauge bosons'' are very different things. A gauge field is a
mathematical symbol that we used to write done a Lagrangian.  While a gauge
boson corresponds to a physical quasiparticle at low energies.
A gauge field in a Lagrangian may not give rise to a gauge boson that appears
as a low energy quasiparticle. Only when the dynamics of gauge field is such
that the gauge field is in the deconfined phase can the gauge boson appears as
a low energy quasiparticle. Thus many researches after the initial finding of
\Ref{BA8880,AM8874} have been concentrated on finding the deconfined phase of
the gauge field.

\begin{figure}
\centerline{
\includegraphics[width=2.3in]{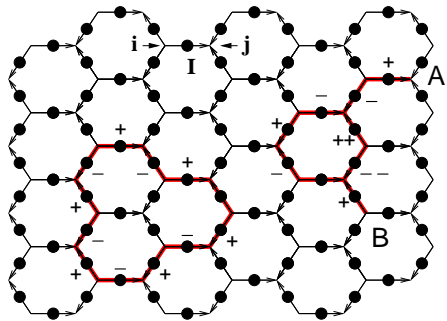}
}
\caption{
The K-lattice formed by filled dots and the H-lattice formed by vertices.  The
spins in our model are on the filled dots.  The lattice contains a closed
string created by alternatively increase and decrease the $\si^z S^z$ of the
spins along a closed path.  The lattice also contains a string-net with nodes
and ends. The string-net is created by
alternatively increase and decrease the $\si^z S^z$ by $1$ or $2$ along the
string-net.  Artificial light corresponds to fluctuations of string-net.  A
pair of artificial charges, A and B,  corresponds to the ends of an open
string.  Note the artificial charges live on the H-lattice.
}
\label{latt}
\end{figure}

Following \Ref{AM8874},
a gapless deconfined $U(1)$ gauge boson was found in a $SU(N)$ spin
model on 3D cubic lattice.\cite{Wlight} The model also contains gapless
fermions.  In 1+3D, the two kinds of excitations can be separated since they
interact weakly at low energies.  We will call those 1+3D excitations,
artificial light and artificial electron.  It was stressed that
the gapless properties of those collective excitations are protected by the
quantum order in the spin ground state.\cite{Wqoslpub,Wlight,WZqoind}
Recently, a simpler and more realistic 3D interacting boson model was found to
contain an artificial light (but not massless fermions).\cite{MS0204}
Exact soluble models and realistic Josephson junction arrays that realize 1+2D
$Z_2$ gauge excitations and related topological order\cite{Wsrvb} can be found
in \Ref{K9721,MS0181,Wqoexct,IFI0203,IF0286,MSP0228} We see that gauge bosons
appear naturally and commonly in quantum ordered states.  We do not need to
introduce them by hand as elementary particles.

Motivated by lattice gauge theory\cite{KS7595} and projection by energy gap
introduced in \Ref{BFG0212,SM0204}, in this paper, we will construct realistic
2D and 3D spin models with screened dipole interaction.  Our models contain an
artificial light as their low energy excitation.  Concrete devices that
realize our models are also designed. Building those devices and observing
artificial light in those devices will show for the first time that elementary
particle, such as light, can be created artificially with designed properties
(such as designed ``speed of light'' and designed value of ``fine structure
constant").

Through our models, we also find that a $U(1)$ gauge theory is actually a
dynamical theory of nets of closed strings.\cite{KS7595} The latter will be
called the string-net theory whose definition will given in section
\ref{string}.  In other words, gauge theory and string-net theory are dual to
each other.  This duality is directly connected to the duality between
statistical $U(1)$ lattice gauge models and statistical membrane
models.\cite{BMK7793,S8053,ID89} According to the string-net picture, a
gapless gauge boson is a fluctuation of large string-nets and charge is the
end of open strings.

In the next a few sections, we will discuss in detail 2D and 3D spin models
and derive their low energy effective theory. For persons who are interested
experimental realization of the spin models and experimental probe of
artificial light, they can go directly to section \ref{sec:real} and
\ref{sec:phy}.

\section{A 2D model}

To construct a realistic model that contains artificial light as its low energy
collective excitation, we consider systems formed by integral spins. We will
consider two cases. In the first case the spins $\v S$ carry magnetic dipole
moment $\v m \propto \v S$. In the second case, we want the spins to carry
electric dipole moment $\v d \propto \v S$.  However, due to the time reversal
symmetry in real molecules, it is impossible for a molecules with a finite
spin to carry an electric dipole moment proportional to the spin. But it is
possible to have a molecule whose ground states are formed by \emph{two} spin
$S$ multiples: $|m,\si^z\>$, with $m=-S,...,+S$ and $\si^z=\pm 1$. We will
call $\si^z$ the $z$-component of isospin. Such a molecule can be viewed as
carrying spin-$S$ and isospin-$1/2$. This kind of molecules can carry a finite
electric dipole moment $\v d\propto \si^z\v S$ and have a time reversal
symmetry, since under time reversal $(\si^z,\v S) \to (-\si^z,-\v S)$.  We
will use the above magnetic dipoles $\v m \propto \v S$ or electric dipoles
$\v d\propto \si^z\v S$ to build our systems.

We start with a honeycomb lattice which will be called the H-lattice. To form
a magnetic dipole system, we place an integral spin-$S$ on every link of the
H-lattice.  For an electric dipole system, we place an integral spin-$S$ and
an isospin-$1/2$ on every link.  We note that the spins form a Kagome lattice
which will be called the K-lattice.  There are two ways to label a spin. We
can use a site index $\v I$ of the K-lattice or we can use a pair of site
indices $\<\v i\v j\>$ that labels a link in the H-lattice (see Fig.
\ref{latt}). Using these two labels, our model Hamiltonian can be written as
\begin{align}
\label{model}
 H & = U \sum_{\v i} 
\left(\sum_{\v\al}
 \si^z_{\<\v i,\v i+\v\al\>} S_{\<\v i,\v i+\v\al\>}^z \right)^2 
+ J \sum_{\v I} (S_{\v I}^z)^2
\nonumber\\
& - \frac{1}{S^2}\sum_{\<\v I\v J\>}\si^z_{\v I}\si^z_{\v J} \left(
t S_{\v I}^{+}S_{\v J}^{-}
+t' S_{\v I}^{+} S_{\v J}^{+}e^{i2\phi_{\v I\v J}} + h.c.
\right)
\end{align}
Here $\v\al$ is one of the three vectors that connect a H-lattice site $\v i$
to its three nearest neighbors. The $U$-term enforces a constraint that the
total $S^z$ of the three spins around a site in the H-lattice is zero.
Also $\phi_{\v I\v J}$ is the angle of the link $\v I\v
J$ in the $xy$-plane and $S^{\pm}=S^x \pm iS^y$.  The summation $\sum_{\<\v I\v J\>}$ is over all
nearest-neighbors in the K-lattice.  The above Hamiltonian applies to both
magnetic dipole systems and electric dipole systems. For magnetic dipole
systems, we regard $\si^z$ as a number $\si^z=1$. For electric dipole systems,
we regard $\si^z$ as the $z$-component of the Pauli matrices.

For the time being, we will treat $\si^z_{\v I}$ classically and assume each
$\si^z_{\v I}$ to take a fixed but a random value of $+1$ or $-1$. (For
magnetic dipole systems, we will set all $\si^z_{\v I}=1$.) Let us first
assume $J=t=t'=0$ and $U>0$. In this case the Hamiltonian is formed by
commuting terms which perform local projections. The ground states are highly
degenerate and form a projected space.  One of the ground states is the state
with $\si^z_{\v I}S_{\v I}^z=0$ for every spin.  Other ground states can be
constructed from the first ground state by drawing a loop in the H-lattice and
then alternatively increase or decrease the $\si^z_{\v I}S_{\v I}^z$ for the
spins on the loop by the same amount.  Such a process can be repeated to
construct all the degenerate ground states.  We see that the projected space
has some non-local characters despite that it is obtained via a local
projection.  

Let us introduce string operator which is formed by the product of $S_{\<\v i\v
j\>}^\pm$ operators
\begin{equation}
\label{string}
 U(C) = \prod_{\<\v i\v j\>} 
S_{\<\v i\v j\>}^{\eta_{\<\v i\v j\>}\si^z_{\<\v i\v j\>}}
\end{equation}
where $C$ is a string connecting nearest-neighbor sites in the H-lattice and
the product $\prod_{\<\v i\v j\>}$ is over all  the nearest-neighbor links of
the H-lattice that form the string.  $\eta_{\<\v i\v j\>}=+1$ if the arrow of
the link $\<\v i\v j\>$ points from $\v i$ to $\v j$ and $\eta_{\<\v i\v
j\>}=-1$ if the arrow of the link $\<\v i\v j\>$ points from $\v j$ to $\v i$.
We note that the string operator alternatively increase or decrease the
$\si^z_{\v I}S_{\v I}^z$ along the string.  If all $\si^z_{\<\v i\v j\>}=1$,
the string operator has the following simple form
\begin{equation}
 U(C) = S^+_{\<\v i_1\v i_2\>}S^-_{\<\v i_2\v i_3\>}S^+_{\<\v i_3\v i_4\>} ...
\end{equation}
where the string $C$ is formed by the H-lattice sites $\v i_1,\v i_2,...$.
Using the string operator, we can create all the degenerate ground states by
repeatedly applying closed-string operators to one of the ground states.

We note that the above string operator $U(C)$ can be defined even when the loop
$C$ intersects or overlaps with itself. In fact, those self
intersecting/overlapping loops are more typical configurations of loops.
Such kind of loops looks like nets of closed strings and we will call them
closed string-nets. (Nets with open strings will be called open string-nets.) 
The string operators $U(C)$ will be called string-net operator.
The degenerate ground states are formed by closed string-nets.

If $t,J\neq 0$, then the ground state degeneracy will be lifted.  The $t$-term
will make string-nets to fluctuate and the $J$-term will give strings in
string-net a string tension. As we will see later the closed string-net
fluctuations become $U(1)$ gauge fluctuations.

The degenerate ground states are invariant under local symmetry
transformations generated by
\begin{equation}
\label{gaugegen}
 U(\phi_{\v i}) = e^{i\sum_{\v i} \left(\eta_{\v i}\phi_{\v i} 
\sum_{\v \al} \si^z_{\<\v i,\v i+\v\al\>}S^z_{\<\v i,\v i+\v\al\>}\right)}
\end{equation}
where $\eta_{\v i}=+1$ if the arrows of links $\<\v i,\v i+\v\al\>$ all point
to $\v i$ and $\eta_{\v i}=-1$ if the arrows of links $\<\v i,\v i+\v\al\>$
all point away from $\v i$ (see Fig.  \ref{latt}). The above transformation is
called the gauge transformation.  Thus we can also say that the degenerate
ground states are gauge invariant.

\section{A four-spin system}
\label{sec:4rotor}

In this section, we will start to derive the low energy effective theory of
our model for the case $t,J\neq 0$ and $t'=0$. We assume $t$ and
$J$ to satisfy $t,J\ll U$ and $J>0$.  The ground state will no longer be
degenerate. The low energy excitations are mainly in the projected space. To
understand the low energy dynamics, we assume $S\gg 1$ and use a semiclassical
approach. 

\begin{figure}
\centerline{
\hfil
\includegraphics[height=1.3in]{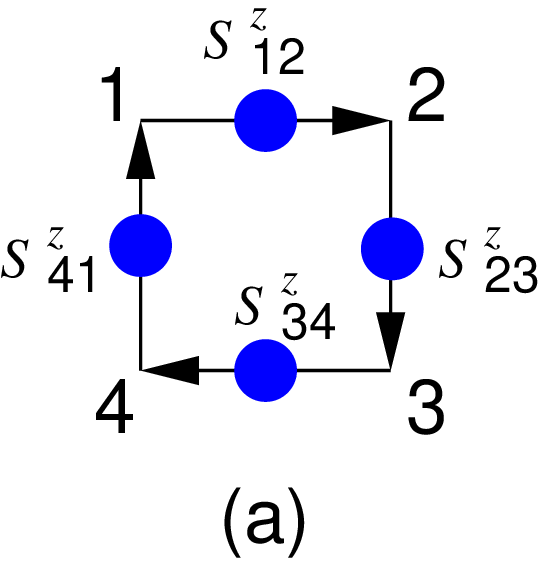}
\hfil
\includegraphics[height=1.2in]{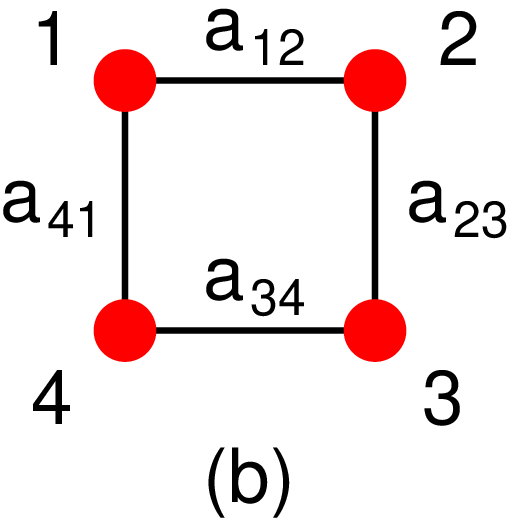}
\hfil
}
\caption{
(a) 
A four-spin system.
(b)
A simple lattice gauge theory is described by lattice gauge field
$a_{i,i+1}$,
and $a_{0,i}$, $i=1,2,3,4$.
}
\label{rotor4}
\end{figure}

To understand the dynamics of our model, let us consider a model of four spins
described by $S^z_{\<12\>}$, $S^z_{\<23\>}$, $S^z_{\<34\>}$, and
$S^z_{\<41\>}$ (see Fig.  \ref{rotor4}a):
\begin{equation}
\label{4Jn2}
H =  \sum_{i}\left( U
(S^z_{\<i-1,i\>} - S^z_{\<i,i+1\>})^2 +
J(S^z_{\<i,i+1\>})^2 \right),
\end{equation}
where we have assumed $4+1\sim 1$ and $1-1 \sim 4$.  The Hilbert space is
spanned by $|n_{\<12\>}n_{\<23\>}n_{\<34\>}n_{\<41\>}\>$ where the integer
$n_{\<i,i+1\>}$ is the eigenvalue of $S^z_{\<i,i+1\>}$. If $U\gg J$, then the
low energy excitations are described by $|nnnn\>$ states with energy $E=4J
n^2$.  All other excitations have energy of order $U$. As we will see in
section \ref{sec:quangauge}, those low energy excitations happen to be
identical to the excitations of a $U(1)$ lattice gauge theory on the same
square (see Fig.  \ref{rotor4}b). Thus our four-spin model describe a gauge
theory at low energies.

To obtain an effective lattice  gauge theory from our spin model, 
we would like to write
down the Lagrangian of our four-spin model.  Since the spins are mainly in the
$x$-$y$ plane, we have $S^{\pm}_{\<ij\>} = S e^{\pm i\th_{\<ij\>}} $.  In this
case $S^z_{\<ij\>}$ is the corresponding momentum $-i\prt/\prt \th_{\<ij\>}$
of variable $\th_{\<ij\>}$.  If we write the Hamiltonian in the form $H=
\frac12 P^T V P$, where $P^T
=(S^z_{\<12\>},S^z_{\<23\>},S^z_{\<34\>},S^z_{\<41\>})$ and 
\begin{equation}
 V = \bpm 
4U+2J & -2U & 0 &-2U \\
-2U   &4U+2J & -2U &0 \\
0 & -2U   &4U+2J & -2U  \\
-2U & 0 & -2U   &4U+2J
\epm  ,
\end{equation}
then the Lagrangian will be 
\begin{equation}
L = \frac12 \dot \Th^T M \dot\Th, 
\end{equation}
where $\Th^T =(\th_{\<12\>},\th_{\<23\>},\th_{\<34\>},\th_{\<41\>})$ and
$M=V^{-1}$.  Obviously, we do not see any sign of gauge theory in the above
Lagrangian. To obtain a gauge theory, we need to derive the Lagrangian in
another way.  Using the path integral representation of $H$, we find
\begin{align}
 Z=&\int D(p)D(\th) e^{i\int dt\left(
\sum_{i} S^z_{\<i,i+1\>}\dot\th_{\<i,i+1\>} - H \right)}
\\
=& \int D(p)D(\th) D(a_0) e^{i\int dt
\left(\sum_i S^z_{\<i,i+1\>}\dot\th_{\<i,i+1\>} - \t H(p,a_0) \right)}  
\nonumber 
\end{align}
where
$
\t H = \sum_{i} \left(J(S^z_{\<i,i+1\>})^2 +  
a_{0,i} (S^z_{\<i-1,i\>}-S^z_{\<i,i+1\>}) 
-\frac{a_{0,i}^2}{4U} \right)
$.
After integrating out  $S^z_{\<i,i+1\>}$, we obtain 
\begin{equation}
Z = \int D(\th)D(a_0) e^{i\int dt L(\th,\dot \th, a_0)}
\end{equation}
where the Lagrangian
\begin{equation}
 L = \frac{1}{4J}\sum_i \left( (\dot \th_{\<i,i+1\>} + a_{0,i}-a_{0,i+1})^2
    +\frac{a_{0,i}^2}{4U} \right)
\end{equation}
In the large $U$ limit, we can drop the $\frac{a_{0,i}^2}{4U}$ term and
obtain
\begin{equation}
 L = \frac{1}{4J}\sum_i (\dot a_{i,i+1} + a_{0,i}-a_{0,i+1})^2
\end{equation}
which is just the Lagrangian of a $U(1)$ lattice gauge theory on a single
square with 
\begin{equation}
a_{i,i+1} = \th_{\<i,i+1\>} , \ \ \ \ a_{i+1,i} =-\th_{\<i,i+1\>}
\end{equation}
as the lattice gauge fields
(see Fig. \ref{rotor4}b).
One can check that the above Lagrangian is invariant under the following 
transformation
\begin{equation}
\label{lattgaugetrans}
 a_{ij}(t) \to a_{ij}(t) + \phi_j(t)-\phi_i(t),\ \ \
 a_{0,i}(t) \to a_{0,i}(t) + \dot \phi_i(t)
\end{equation}
which is called the gauge transformation.

We note that low energy wave function $\Psi(a_{12}, a_{23},a_{34},a_{41})$ is
a superposition of $|nnnn\>$ states. All the low energy states are gauge
invariant, \ie invariant under gauge transformation
$a_{ij} \to a_{ij} + \phi_j-\phi_i$.

The electric field of a continuum $U(1)$ gauge theory is given by
$\v e = \dot{\v a} - \v \prt a_0$. 
In a lattice gauge theory, the electric
field becomes a quantity defined on the links
\begin{equation}
 e_{ij} = \dot a_{ij} - (a_{0,j} - a_{0,i})
\end{equation}
We see that our lattice gauge Lagrangian can be written as
$L = \frac{1}{4J}\sum_i e_{i,i+1}^2$. Comparing with the 
continuum $U(1)$ gauge theory $\cL \propto \v e^2 - \v b^2$, we see that our
Lagrangian contains only the kinetic energy corresponding to $\v e^2$.
A more general lattice gauge theory also contains a potential energy term
corresponding to $\v b^2$.

To obtain a potential energy term, we generalized our spin model to
\begin{align}
\label{HrotorUJt}
H =&  \sum_{i}\big( U
(S^z_{\<i-1,i\>} - S^z_{\<i,i+1\>})^2 +
J(S^z_{\<i,i+1\>})^2 \nonumber\\
& + t(e^{i\th_{\<i-1,i\>}} e^{i\th_{\<i,i+1\>}} +h.c.) \big),
\end{align}
We note that $\<nnn| e^{i\th_{\<i-1,i\>}} e^{-i\th_{\<i,i+1\>}} |nnn\>=0$. Thus
at the first order of $t$, the new term has no effect at low energies.
The low energy effect of new term only appear at the second order of $t$.

We can repeat the above calculation to obtain the following Lagrangian
\begin{align}
 L =& \frac{1}{4J}\sum_i \big((\dot a_{i,i+1} + a_{0,i}-a_{0,i+1})^2
\nonumber\\
   &  - t(e^{i(a_{i-1,i}+a_{i,i+1})} +h.c.) +\frac{a_{0,i}^2}{4U}\big)
\end{align}
It is a little more difficult to see in Lagrangian why the new term has no low
energy effect at the first order of $t$. 
Let us concentrate on the fluctuations of the following form
\begin{equation}
a_{i-1,i} =\phi_i-\phi_{i-1}
\end{equation}
In the lattice gauge theory, such type of fluctuations are called the pure
gauge fluctuations.  After integrating out $a_{0,i}$, the Lagrangian for the
above type of fluctuations has a form $L= \frac12 \dot \phi_i m_{ij} \dot
\phi_j - \sum_i t( e^{i(\phi_{i-1}-\phi_{i+1})} +h.c.)$ with
$m_{ij}=O(U^{-1})$.  We see that, in the large $U$ limit, the above form of
fluctuations are fast fluctuations. Since $\phi_i$ live on a compact space
(\ie $\phi_i$ and $\phi_i+2\pi$ represent the same point), those fast
fluctuations all have large energy gap of order $U$. Now we see that the
$t$-term $t e^{i(\phi_{i-1}-\phi_{i+1})}$ average to zero for fast
fluctuations and has no effect at the first order in $t$. However, at second
order in $t$ there is a term 
$t^2 \prod_{i=1,3} e^{i(\th_{\<i-1,i\>}+\th_{\<i,i+1\>})} = t^2
e^{i \sum_i \th_{\<i-1,i\>}}$. Such a term does not depend on $\phi_i$ and does
not average to zero. Thus we expect the low energy effective Lagrangian to
have a form
\begin{equation}
\label{gauge4U}
 L = \frac{1}{4J}\sum_i \left((\dot a_{i,i+1} + a_{0,i}-a_{0,i+1})^2
    +\frac{a_{0,i}^2}{4U}\right) + g \cos\Phi
\end{equation}
where $g=O(t^2/U)$ and $\Phi = \sum_i a_{i,i+1}$ is the flux of the $U(1)$
gauge field through the square.

To calculate $g$ quantitatively, we would like to first derive the low energy
effective Hamiltonian. If we treat the $t$-term as a perturbation and treat
the low energy states as degenerate states, then at second order in $t$, we
have
\begin{align}
& <n',n',n',n'| H_{eff} |nnnn\>  \nonumber\\
=& -\frac{t^2}{2U} 
\<n',n',n',n')|e^{i(\th_{\<34\>}+\th_{\<41\>})}|n',n',n,n\> \times \nonumber\\
&\ \<n',n',n,n|  e^{i(\th_{\<12\>}+\th_{\<23\>})} |nnnn\> \nonumber\\
& +\hbox{three other similar terms} \nonumber\\
=& -\frac{2t^2}{U}
\end{align}
where $n'=n+1$.
Thus the low energy effective Hamiltonian is
\begin{equation}
\sum_{i}\left( U
(S^z_{\<i-1,i\>} - S^z_{\<i,i+1\>})^2 +
J(S^z_{\<i,i+1\>})^2 \right) -\frac{4t^2}{U} \cos(\Phi) .
\end{equation}
The corresponding Lagrangian is given by \Eq{gauge4U} with 
\begin{equation}
 g = \frac{4t^2}{U}.
\end{equation}
As discussed before, the pure gauge fluctuations has a large energy gap of
order $U$. The low energy effective theory below $U$ can be obtained by letting
$U\to \infty$ and we get
\begin{equation}
\label{gauge4}
 L = \frac{1}{4J}\sum_i (\dot a_{i,i+1}+a_{0,i}-a_{0,i+1})^2+g\cos\Phi
\end{equation}
which contains both electric energy and magnetic energy.

\section{Quantum gauge theory}
\label{sec:quangauge}

In this section, we will reverse the above calculation and start with the
classical lattice gauge theory described by the Lagrangian \Eq{gauge4}. We
would like to quantize it and find its Hamiltonian.  This will allow us to
calculate the energy levels of the lattice gauge theory and compare them with
the energy levels of the four-spin model.

As a gauge theory, the path integral
\begin{equation}
 Z=\int D(a)D(a_0) e^{-i \int dt \left(
\frac{1}{4J}\sum_i (\dot a_{i,i+1}+a_{0,i}-a_{0,i+1})^2+g\cos\Phi
\right)} ,
\end{equation}
should not be regarded as a summation over different functions $(a_{ij}(t),
a_{0,i}(t))$. Here we regard two paths related by the gauge transformation
\Eq{lattgaugetrans} as the same path. Thus the path integral should be
regarded as a summation over gauge equivalent classes of paths.  Thus
$(a_{ij}(t),a_{0,i}(t))$ is a many-to-one label of the gauge equivalent
classes. We can obtain a one-to-one label by ``fixing a gauge''.  We note that
$\sum_j a_{ij}$ transforms as $\sum_j a_{ij}\to \sum_j \t a_{ij}= \sum_j
(a_{ij} +\phi_i-\phi_j)$ under gauge transformation. By tuning $\phi_i$, we
can always make $\sum_j \t a_{ij}=0$. Thus for any path
$(a_{ij}(t),a_{0,i}(t))$, we can always make a gauge transformation to make
$\sum_j a_{ij}=0$. Therefore, we can fix a gauge by choosing a gauge fixing
condition
\begin{equation}
\sum_j a_{ij}=0
\end{equation}
Such a gauge is called the Coulomb gauge, which has a form
$\v\prt\cdot \v a=0$ for a continuum theory.
In the Coulomb gauge our path integral becomes
\begin{align}
 Z=&\int D(a) D(a_0) \prod_i \del( \sum_j a_{ij}) \nonumber\\ 
& \ \ e^{-i \int dt \left(
\frac{1}{4J}\sum_i (\dot a_{i,i+1}+a_{0,i}-a_{0,i+1})^2+g\cos\Phi
\right)}
\end{align}
We note that a coupling between $a_{0,i}$ and $a_{ij}$ has a form
$a_{0,i} \sum_j \dot a_{ij}$. Thus
for $a_{ij}$ satisfying the constraint $\sum_j a_{ij}=0$, $a_{0,i}$ and
$a_{ij}$ do not couple. Since $a_{0,i}$ has no dynamics (\ie no $\dot a_{0,i}$
terms), we can integrate out
$a_{0,i}$. The resulting path integral becomes
\begin{equation}
 Z=\int D(a) \prod_i \del( \sum_j a_{ij}) 
e^{-i \int dt \left(
\frac{1}{4J}\sum_i \dot a_{i,i+1}^2+g\cos\Phi
\right)}
\end{equation}
which is the path integral in the Coulomb gauge.

In general, a path integral in the Coulomb gauge can be obtained by the
following two simple steps: (a) inserting the gauge fixing condition $\prod_i
\del( \sum_j a_{ij})$ and (b) drop the $a_{0,i}$ field.

For our problem, the constraint $\prod_i \del( \sum_j a_{ij})$ makes
$a_{12}=a_{23}=a_{34}=a_{41}\equiv \th/4$. The path integral takes a simple 
form
\begin{equation}
\label{thcosth}
 Z=\int D(\th) e^{-i \int dt \left( \frac{1}{16J} \dot \th^2+g\cos \th
\right)}
\end{equation}
we note that the configuration
$(a_{12},a_{23},a_{34},a_{41})=(\pi/2,\pi/2,\pi/2,\pi/2)$ is gauge equivalent
to $(a_{12},a_{23},a_{34},a_{41})=(2\pi,0,0,0)$ (\ie there is a gauge
transformation that transform $(\pi/2,\pi/2,\pi/2,\pi/2)$ to $(2\pi,0,0,0)$).
Also $a_{12}=2\pi$ is equivalent to $a_{12}=0$ since $a_{i,i+1}=\th_{\<i,i+1\>}$
live on a circle. Thus $\th=2\pi$ and $\th=0$ correspond to the same physical
point.  The path integral \Eq{thcosth} describes a particle of mass
$(8J)^{-1}$ on a unit circle. The flux energy $-g\cos \th$ is the potential
experienced by the particle. When $g=0$, the energy levels are given by
$E_n=4Jn^2$ which agrees exactly with the energy levels of \Eq{4Jn2} at low
energies. Hence \Eq{4Jn2} is indeed a gauge theory at low energies.

\section{Effective gauge theory of lattice spin model}

Using the similar calculation, we find that our 2D lattice model \Eq{model}
can be described by the
following Lagrangian in the large $U$ limit
\begin{align}
\label{lgauge}
 L = &
 \frac{1}{4\t J}\sum_{\<\v i\v j\>}
[\dot a_{\v i\v j} + a_0(\v i)-a_0(\v j)]^2 
\nonumber\\
&+ g \sum_{\v p} \eta_{\v p} \cos(\Phi_{\v p})
+ J_1 \sum_{\<\v I\v J\>} \si^z_{\v I}\si^z_{\v J} 
\end{align}
Here $a_{\v i\v j} = \th_{\<\v i\v j\>}$ if the arrow of link (see Fig.
\ref{latt}) points from $\v i$ to $\v j$ and $a_{\v i\v j} = -\th_{\<\v i\v
j\>}$ if the arrow points from $\v j$ to $\v i$. $\v p$ labels the plaquettes
in the H-lattice and $\Phi_{\v p} = a_{12}+a_{23}+...+a_{61}$, where 1,...,6
are the six sites around the plaquette $\v p$. The $ \sum_{\<\v I\v J\>}$
sums over all the nearest neighbor sites $\<\v I\v J\>$ in the K-lattice.
$\eta_{\v p}=1$ if all
$\si^z_{12}$, ..., $\si^z_{61}$ are equal and $0\leq \eta_{\v p} \alt 0.5$ 
otherwise.  In the small $t$ limit $\t J=J$.  

Let us first explain the potential term $-J_1 \sum_{\<\v I\v J\>} \si^z_{\v
I}\si^z_{\v J}$.  We start with a low energy state in the projected space
$|\Psi\>$.  The action of the $t$-term $tS^{-2} \si^z_{\v I}\si^z_{\v J}
S^+_{\v I} S^-_{\v J}$ on such a state give us a high energy state with an
energy $4U - 2\si^z_{\v I}\si^z_{\v J} U$. The second order perturbation in
$t$ gives rise to the following contribution 
$-2\times t^2S^{-4}/(4U - 2\si^z_{\v I}\si^z_{\v J} U)$. 
We see that $\si^z_{\v I}\si^z_{\v J}=1$ has a lower
energy than $\si^z_{\v I}\si^z_{\v J}=-1$. The energy difference is
$2t^2S^{-4}/3U$.  We find that $J_1 = t^2S^{-4}/3U$. The dynamics of the
isospin $\si^z$ is described by an Ising model. The ground state is a
ferromagnetic state with all $\si^z_{\v I}=1$ (or $\si^z_{\v I}=-1$).

At second order, the $t$-term can also generate the
$J$-term in \Eq{model}. Thus $\t J-J \sim t^2/U$.  

Second,
let us explain the potential term $-g \sum_{\v p} \eta_{\v p} \cos(\Phi_{\v
p})$.  
We first note that the gauge transformation \Eq{gaugegen} changes
\begin{equation} 
\label{gaugetran} 
a_{\v i\v j} \to a_{\v i\v j}+\phi_{\v
i}-\phi_{\v j}.  
\end{equation} 
The $t$-term in \Eq{model} can be written as $S^2\cos(\si^z_{\<\v i\v
j\>}a_{\v i\v j}+\si^z_{\<\v j\v k\>}a_{\v j\v k})$ which is not gauge
invariant.  Thus the average of the $t$-term in the projected space is zero.
Non zero potential terms can only be generated from the $t$-term via higher
order perturbation, and the resulting potential term must be gauge invariant.
The simplest gauge invariant term has a form $\cos(\Phi_{\v p})$ which is
generated at the third order in $t/U$.
Hence $g \sim t^3S^{-6}/U^2$. In the small $t$ limit, the second order $J_1$
term will make all $\si^z_{\v I}=1$. In the following we will calculate the
$g$-term assuming  $\si^z_{\v I}=1$.

At third order, the effective Hamiltonian in the projected space has the
following matrix elements
\begin{align}
 \<\Psi_1|H_{eff}|\Psi_2\> =&
{\sum_{m,n}}^\prime \frac{\<\Psi_1| H_t |m\>\<m| H_t|n\>\<n|H_t |\Psi_2\>}{
(E_m-E_\Psi)(E_n-E_\Psi)}
\end{align}
where $E_\Psi\sim 0$ is the energy of $|\Psi_{1,2}\>$, $\sum^\prime$ is a sum
over all high energy states $|m\>$ and $|n\>$ that are \emph{not} in the
projected space, and $H_t$ is the $t$-term $H_t=t S^{-2} \sum_{\<\v I\v J\>}
\si^z_{\v I}\si^z_{\v J} S^+_{\v I} S^-_{\v J}$.  When $|\Psi_{1}\>=
e^{i\Phi_{\v p}}|\Psi_{1,2}\>$, 
we find $\<\Psi_1|H_{eff}|\Psi_2\>= 6\times 2\times t^3S^{-6}/(2U)^2$.  Thus
$g\eta_{\v p}=6t^3S^{-6}/U^2$.
 
In a numerical calculation, we considered our model on a single hexagon - a
single cell of the H-lattice and assumed $S=1$.  Solving the six-spin model
exactly, we found that the low energy sector and the high energy sector start
to mix when $g\sim 0.25 U$. In that case perturbation theory break down.

The $J_1$-term favors a ground state with all
$\si^z_{\v I}=1$ or $\si^z_{\v I}=-1$. Such a ground state spontaneously
breaks the time reversal symmetry. The time reversal symmetry breaking happens
even when we include the quantum fluctuations of $\si^z_{\v I}$ generated by
$\del H = J'\sum_{\v I} \si^x_{\v I}$ as long as $J'\alt {\rm max}(|g|,
t^2/U)$. In the time reversal symmetry breaking phase, \Eq{lgauge} describes a
$U(1)$ lattice gauge theory.

When $t'\neq 0$, more complicated term of form $\cos(\Phi_{\v p} +\phi)$ can
be generated, where $\phi$ depends on $\si^z_{12}$,...,$\si^z_{61}$. In this
case, $\si^z_{\v I}$ might have a certain pattern in the ground state which
can break translation and/or rotation symmetry. But as long as $J'$ is small,
the quantum fluctuations of $\si^z_{\v I}$ can be ignored and the model
contains a $U(1)$ gapless gauge boson if we ignore the instanton effect.

However, in 2+1D, we do have an instanton effect.  Due to the instanton
effect, a $U(1)$ gauge excitation develops a gap.\cite{P7729} The instanton
effect is associated with a change of the $U(1)$ flux $\Phi$ from $0$ to
$2\pi$ on a plaquette.  To estimate the importance of the instanton effect,
let us consider a model with only a single plaquette (\ie the single-hexagon
model discussed before).  Such a model is described by
\begin{equation}
L = \frac{1}{24\t J} \dot \th^2+g\cos \th .
\end{equation}
The instanton effect corresponds to a path $\th(t)$
where $\th$ goes from $\th(-\infty)=0$ to $\th(+\infty)=2\pi$.  To estimate
the instanton action, we assume
\begin{equation}
 \th(t) =
\begin{cases} \displaystyle\
0, & \hbox{for } t<0 \\
\displaystyle\
2\pi t/T, & \hbox{for } 0< t<T \\
\displaystyle\
2\pi, & \hbox{for } T< t
\end{cases} 
\end{equation}
The minimal instanton action is found to be
\begin{equation}
 S_c = \pi \sqrt{2g/3\t J}
\end{equation}
when $T=\pi/2\sqrt{3g\t J/2}$.
From the density of the instanton gas $\sqrt{\t J g}e^{-S_c}$, we estimate the energy gap
of the $U(1)$ gauge boson to be
\begin{equation}
\label{lightgap}
 \Del \sim \sqrt{\t J g} e^{- \pi \sqrt{2g/3\t J}}
\end{equation}
Thus to have a
nearly gapless gauge boson, we require the above gap to be much less than the
bandwidth of the gauge field $\sqrt{g\t J}$.  
This requires
\begin{equation}
\label{cond}
g \alt 0.25U, \ \ \ \ e^{-2.4\sqrt{\frac{g}{\t J}}} \ll 1
\end{equation}
If the above condition is satisfied, we can ignore the mass gap of the gauge
boson and regard the $U(1)$ gauge theory as in the deconfined phase.
Therefore, \Eq{cond} is the conditions to have an artificial light in our 2D
model.

\section{String-net theory and string-net picture of 
artificial light and artificial charge}

As mention before, the  low energy excitations below $U$ are describe by
closed string-nets of increased/decreased $\si^z S^z$.
(see Fig.  \ref{latt}).  To
make this picture more precise, we would like to define a 
closed-string-net theory on a lattice. 

The Hilbert space of the closed-string-net theory is a subspace of the Hilbert
space of our model \Eq{model} (here we assume all $\si^z_{\v I}=1$). The
closed-string-net Hilbert space contains a state with all $S^z_{\v I}=0$.  If
we apply the closed-string-net operator \Eq{string} to the $S^z_{\v I}=0$
state, we obtain another state in the closed-string-net Hilbert space. Such a
state is formed by $S^z_{\v I}=\pm 1$ along the closed loop, or more generally
a closed string-net $C$ if we include self intersection and overlap. Thus
$U(C)$ in \Eq{string} can be viewed as a string-net creation operator.  Other
states in the closed-string-net Hilbert space correspond to
multiple-string-net states and are generated by repeatedly applying the
closed-string-net operators \Eq{string} to the $S^z_{\v I}=0$ state.  

The Hamiltonian of our closed-string-net theory is given by
\begin{equation}
\label{strH}
 H_{str} = \sum_{\v I} \t J (S^z_{\v I})^2 
- \sum_{\v p} \frac12  ( g W_{\v p}+h.c.)
\end{equation}
where $\sum_{\v p}$ sums over all the plaquettes of the H-lattice, and $W_{\v
p}$ is the closed-string-net operator for the closed string around the
plaquette $\v p$.  One can check that the above Hamiltonian acts within the
closed-string-net Hilbert space.  The $\t J$ term gives strings in string-nets
a finite string tension, and the $g$ term causes the string-nets to fluctuate.

From the construction, it is clear that the closed-string-net Hilbert space is
identical to the low energy Hilbert space of our model \Eq{model} which is
formed by states with energy less than $U$. From our derivation of effective
lattice gauge theory \Eq{lgauge}, it is also clear that the closed-string-net
Hamiltonian \Eq{strH} is directly related to the lattice gauge Lagrangian
\Eq{lgauge}.  In fact, the Hamiltonian of the lattice gauge theory is
identical to the closed-string-net Hamiltonian \Eq{strH}.  The string tension
$\sum_{\v I} \t J (S^z_{\v I})^2$ term in the string-net theory corresponds to
the $\frac{1}{4\t J}\sum_{\<\v i\v j\>} [\dot a_{\v i\v j} + a_0(\v i)-a_0(\v
j)]^2 $ term in the gauge theory, and the string hopping $\sum_{\v p} \frac12
g (W_{\v p}+h.c.)$ term in the string-net theory corresponds to the $g
\sum_{\v p} \eta_{\v p} \cos(\Phi_{\v p})$ term in the gauge theory.  Since
the $S^z\sim \dot \th_{\<\v i\v j\>}= \dot a_{\v i\v j}$ corresponds to the
electric flux along the link, A closed loop of increased/decreased $\si^z S^z$
corresponds to a loop of electric flux tube. A string-net corresponds a
``river'' network of electric flux.  

We see that the $U(1)$ gauge theory \Eq{lgauge} is actually a dynamical theory
of nets of closed strings. Typically, one expects a dynamical theory of
closed-string-nets to be written in terms of string-nets as in \Eq{strH}.
However, since we are more familiar with field theory, what we did in the last
a few sections can be viewed as an attempt trying to describe a string-net
theory using a field theory.  Through some mathematical trick, we have
achieved our goal. We are able to write the string-net theory in a form of
gauge field theory. The gauge field theory is a special field theory in which
the field \emph{does not} correspond to physical degrees of freedom and the
physical Hilbert space is non-local (in the sense that the total physical
Hilbert space cannot be written as a direct product of local Hilbert spaces).
The point we try to make here is that gauge theory (at least the one discussed
here) is a closed-string-net theory in disguise. Or in other words, gauge
theory and closed-string-net theory are dual to each other.  We would like to
point out that in \Ref{S8053,ID89} various duality relations between lattice
gauge theories and theories of extended objects were reviewed. In particular,
some statistical lattice gauge models were found to be dual to certain
statistical membrane models.\cite{BMK7793} This duality relation is directly
connected to the relation between gauge theory and closed-string-net theory in
our dipole models.

In the large $\t J/g$ (hence large $\Del_{gauge}$) limit, the ground states
for both the dipole model and string-net model are
given by $S^z=0$ for every spin. In this phase, the closed string-nets or the
electric flux tubes do not fluctuate much and have an energy proportional to
their length. This implies that the $U(1)$ gauge theory is in the confining
phase.  In the small $\t J/g$ limit, 
the closed string-nets fluctuate strongly and the space is filled with closed
string-nets of arbitrary sizes. According to the calculation in the previous
section, we note that the small $\t J/g$ phase can also be viewed as the
Coulomb phase with gapless gauge bosons. Combining the two pictures, we see
that gapless gauge bosons correspond to fluctuations of large closed
string-nets. 

After relating the closed strings (or closed string-nets) to artificial light,
we now turn to artificial charges.  To create a pair of particles with
opposite artificial charges for the artificial $U(1)$ gauge field, we need to
draw an open string (or an open string-net) 
and alternatively increase and decrease the $\si^z S^z$ of
the spins along the string (see Fig.  \ref{latt}).  
The end points of the open
strings, as the end points of electric flux tubes, correspond to particles
with opposite artificial charges. We note that charged particles live on the
H-lattice.  In the confining phase, the string connecting the two artificial
charges does not fluctuate much. The energy of the string is proportional to
the length of the string. We see that there is a linear confinement between
the artificial charges.  

In the small $\t J/g$ limit, the large $g$ cause strong fluctuations of the
closed string-nets, which lead to gapless $U(1)$ gauge fluctuations.  The
strong fluctuations of the string connecting the two charges also changes the
linear confining potential to the $\log(r)$ potential between the charges. 

To understand the dynamics of particles with artificial charges, let us
derive the low energy effective theory for those charged particles.
Let us first assume $J=t=t'=0$. A pair of charged particles with opposite 
unit artificial charges can be created by applying the open-string operator
\Eq{string} to the ground state. We find that each charge particle has a
energy $U$ and the string costs no energy. Let us first treat charge particles
as independent particles. In this case the total
Hilbert space of charged particles is formed by state
$|\{n_{\v i}\}\>$, where $n_{\v i}$ is the number of artificial charges on the
site $\v i$ of the H-lattice. $|\{n_{\v i}\}\>$ is an energy eigenstate with
energy $E = U\sum_{\v i} n_{\v i}^2$. Such a system can be described by the
following rotor Lagrangian
\begin{equation}
 L = \sum_{\v i} \frac{1}{4U} \dot \vphi_{\v i}^2
\end{equation}
where $\vphi_{\v i}$ is an angular variable.  The creation operator of the
charged particle is given by $e^{i\vphi_{\v i}}$.  Now, let us include the
fact that the charged particles are always the ends of open strings (or nodes
of string-nets).  Such a fact can be implemented by including the $U(1)$ gauge
field in the above Lagrangian. Using the gauge invariance, we find the gauged
Lagrangian has a form
\begin{equation}
 L = \sum_{\v i} \frac{1}{4U} (\dot \vphi_{\v i} + a_0(\v i))^2
\end{equation}
After including the gauge field, the single charge creation operator
$e^{i\vphi_{\v i}}$ is no longer physical since it is not gauge invariant. The
gauge invariant operator 
\begin{equation}
e^{-i\vphi_{\v i_1}} e^{ia_{\v i_1\v i_2}} ... e^{ia_{\v i_{N-1}\v i_N}} 
e^{i\vphi_{\v i_N}}
\end{equation}
always creates a pair of opposite charges. In fact the above gauge invariant
operator is nothing but the open-string-net operator \Eq{string}.  We also see
that the string-net operator \Eq{string} is closely related to the 
Wegner-Wilson loop operator.\cite{W7159,W7445,K7959}

The $t$-term generates a hopping of charged particles to the next-nearest
neighbor in the H-lattice. Thus, if $t\neq 0$, the charged particles will have
a non-trivial dispersion. The corresponding Lagrangian is given by
\begin{equation}
\label{Lcharged}
 L = \sum_{\v i} \frac{(\dot \vphi_{\v i} + a_0(\v i))^2}{4U} 
- \sum_{(\v i\v j)} t( 
e^{ i(\vphi_{\v i} -\vphi_{\v j} - a_{\v i\v k} - a_{\v k\v j})} + h.c.)
\end{equation}
where $(\v i\v j)$ are next-nearest neighbors in the H-lattice, and $\v k$ is
the site between site $\v i$ and site $\v j$.  The above Lagrangian also tells
us that the charged particles are bosons.  We also note that a flipped spin
corresponds to two artificial charges.  Therefore each unit of artificial
charge corresponds to a half-integer spin.

Using the string-net picture, we can give more concrete answers to the three
questions about light:  \\
\fbox{\parbox{3.3in}{
What is light? \\
Light is a fluctuation of closed string-nets of arbitrary sizes.}}
\fbox{\parbox{3.3in}{
Where light comes from?  \\
Light comes from the collective motions of ``things'' that our vacuum is made 
of. \footnote{Note that our vacuum is not empty. It is filled with ``things'' that form 
the space-time.} In particular, light come from the large closed string-nets 
that fill the vacuum. }}
\fbox{\parbox{3.3in}{
Why light exists?   \\
Light exists because our vacuum contains strong
fluctuations of loop-like objects (the closed string-nets) of 
arbitrary size. }}

We would like to stress that the above string-net picture of the actual light
in nature is just a proposal. There may be other theories that explain what is
light and where light comes from. In this paper, we try to argue that the
string-net picture is at least self consistent, since there are actual models
that realize the string-net picture of light. We also try to argue that the
string-net picture of light is more natural than the current theory of light
where light is regarded as a vector gauge field which is introduced by hand.

\begin{figure}
\centerline{
\includegraphics[width=1.5in]{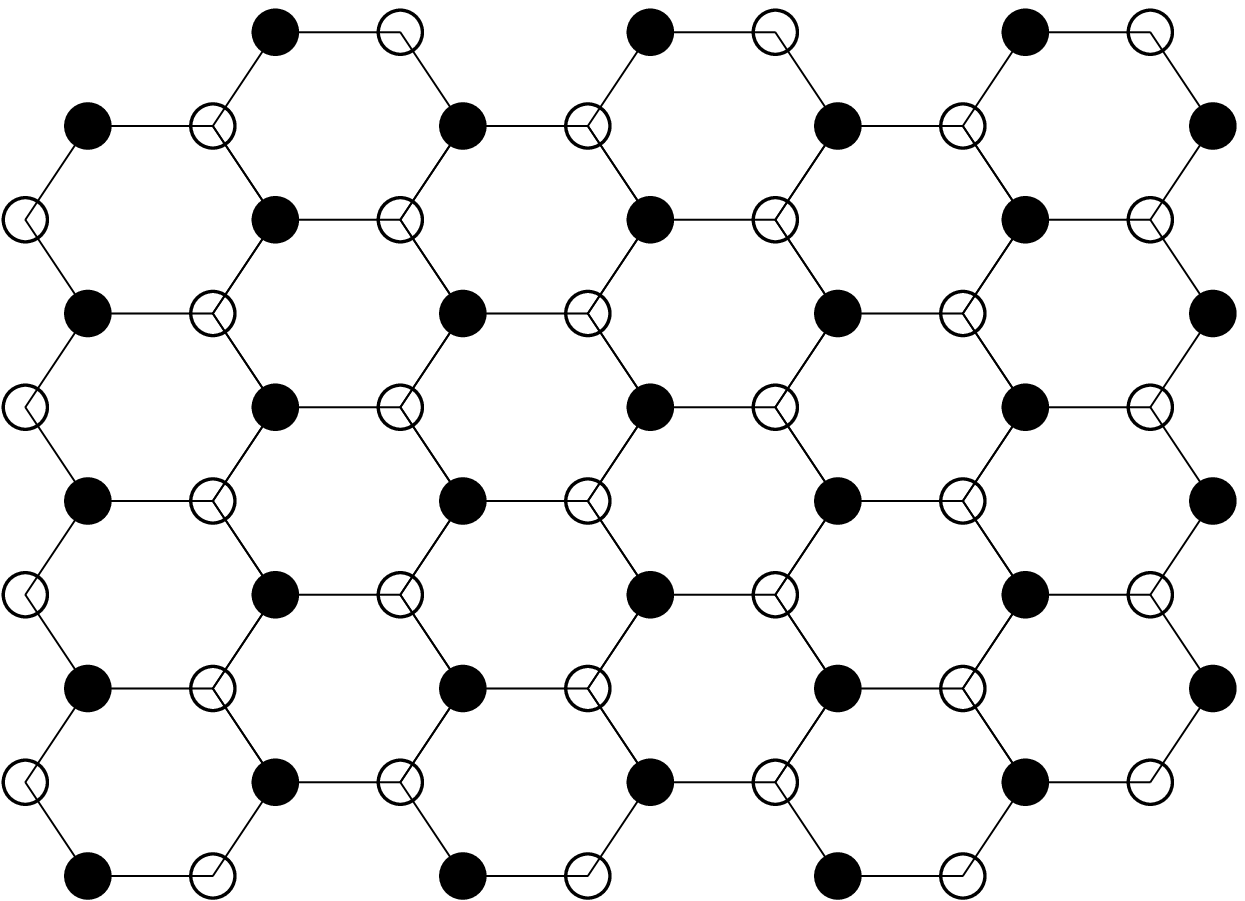}
}
\caption{
Two sublattices of a H-lattice. In a 3DH-lattice, the open dots link to the
layer above and the filled dots to the layer below.
}
\label{latt3D}
\end{figure}

\section{ A 3D model}

Our 2D model and the related calculations can be easily generalized to 3
dimensions. To construct our 3D model, we first construct a 3DH-lattice which
is formed by layers of H-lattices stacked on top of each other. Note that a
H-lattice can be divided into two triangular sublattices (see Fig.
\ref{latt3D}).  We link the sites in one sublattice to the corresponding sites
in the layer above and link the sites in the other sublattice to the layer
below. The spins are place on the links of the 3DH-lattice. The lattice formed
by the spins is called 3DK-lattice. Actually, the 3DK-lattice is nothing but
the conner-sharing tetrahedron lattice or the pyrochlore lattice.  The 3D
Hamiltonian still has a form \Eq{model}. But now $\v i$ label the sites in the
3DH-lattice and $\v I$ the sites in the 3DK-lattice. $\v \al$ connects the
site $\v i$ to its four linked neighbors in the 3DH-lattice.  The low energy
effective theory still has the form \Eq{lgauge} and the conditions to observe
artificial light are still given by \Eq{cond}.  The main difference between
the 1+2D model and 1+3D model is that the artificial light, if exist, is
exactly gapless in 1+3D.  The effective Lagrangian for the charged particles
still has the form in \Eq{Lcharged}.

\section{Emerging quantum order}

Our 3D model contains two $T=0$ quantum phases with the same symmetry. One
phase (phase A) appears in $\t J\gg |g|$ limit and is gapped (see \Eq{cond}).
The other phase (phase B) appears in $|g|\gg \t J$ limit.  The phase B
contains a non-trivial quantum order which is closely related to the
artificial light in it.  

By including the $t$-term between spins beyond nearest neighbors, our model
can even support different kinds of non-trivial quantum orders.  For example,
by adjusting the different $t$-terms, we can independently tune the value and
the sign of $g$ in $g\cos(\Phi_{\v p})$ for different kind of plaquettes. If
all $g$ are positive, then we get the phase B discussed above, where there is
zero gauge flux through all the plaquettes: $e^{i\Phi_{\v p}}\sim 1$.  If we
tune $g$ to be negative for the plaquettes in the layers of the 3DH-lattice
and positive for the plaquettes between the layers, then we get a phase (phase
C) with a new quantum order. In phase C there is $\pi$ flux through the
plaquettes in the layers and zero flux through the other plaquettes.  The
phase C has the same symmetry as the phase A and B, and contains a gapless
artificial light.  The phase B and phase C are separated by phase A that
appears in small $g$ limit.

Quantum orders in phase B and C can be more precisely characterized by the
projective symmetry group or PSG.\cite{Wqoslpub,Wqogen} In semiclassical
limit, the phase B is described by ansatz where all $\<e^{ia_{\v i\v j}}\>\sim
1$. While the phase C is described by ansatz where some $\<e^{ia_{\v i\v
j}}\>\sim 1$ and other $\<e^{ia_{\v i\v j}}\>\sim -1$. The PSG for an ansatz
is formed by all the combined gauge and symmetry transformations that leave
the ansatz invariant.\cite{Wqoslpub,Wqogen}  We find the PSG's for the ansatz
of phase B and the ansatz of phase C are different.  It was shown that PSG is
a universal property of a quantum phase that can be changed only through phase
transitions.\cite{Wqoslpub,Wqogen} The different PSG's for the phase B and
phase C indicate that phase B and phase C are indeed different quantum phases
which cannot be changed into each other at $T=0$ without a phase transition.
Using PSG we can also describe more complicated quantum orders (or flux
configurations).  We can even use PSG to classify all the quantum orders in
our model (in semiclassical limit).

The different quantum orders in the phase B and phase C can be distinguished in
experiments by measuring the dispersion relation of the charge particle.
From \Eq{Lcharged}, we see that the hoping of the charged particles is
affected by the flux through the plaquettes.

\section{Emerging low energy gauge invariance}

After seeing the importance of gauge transformation \Eq{gaugegen} in obtaining
artificial light and in PSG characterization of quantum orders, we are ready
to make a remark about the gauge invariance.  We note that after including the
higher order $t/U$ terms, the Lagrangian formally is not invariant under gauge
transformation \Eq{gaugetran}. As a result, the so called pure gauge
fluctuations (which should be unphysical in gauge theory) actually represent
physical degrees of freedom. However, those fluctuations 
all have a large energy gap of order $U$.\cite{BFG0212,SM0204} The low energy
fluctuations (assuming there is a finite energy gap between the low energy and
high energy excitations) should be gauge invariant, and the effective Lagrangian
that describes their dynamics should be gauge invariant.  

Due to the finite mixing between the low energy and high energy excitations
caused by the $t$-term, the low energy excitations are not invariant under the
particular gauge transformation defined in \Eq{gaugegen}. However, since the
mixing is perturbative, we can perturbatively modify the gauge transformations
such that the low energy excitations are invariant under a modified gauge
transformation.  To obtain the modified gauge transformation, we continuously
change $t$ from zero to a small value.  This will cause the eigenstates of our
model to rotate.  The rotation is generated by a unitary
matrix $W$. Then the modified gauge transformation is given by $ \t U(\phi_{\v
i})= W e^{i\sum_{\v i} \left(\eta_{\v i}\phi_{\v i} \sum_{\v \al} \si^z_{\<\v
i,\v i+\v\al\>} S^z_{\<\v i,\v i+\v\al\>}\right)} W^\dag $. By definition, the
modified gauge transformation will leave the low energy excitations invariant.
The non-trivial point here is that the modified gauge generator $W \sum_{\v i}
\left(\eta_{\v i}\phi_{\v i} \sum_{\v \al} \si^z_{\<\v i,\v i+\v\al\>}
S^z_{\<\v i,\v i+\v\al\>}\right) W^\dag $ is still a local operator. This is
likely to be the case if $t$ is not too large to destroy the energy gap
between the low and high energy excitations.  We see that both the
$U(1)$ gauge structure and the PSG are emerging properties in our model.

We would like to remark that the key to obtain a low
energy effective gauge theory is not to formally derive an effective
Lagrangian that have a gauge invariance, but to show all the pure gauge
fluctuations to have a large energy gap. In this limit, as we have seen for
the $t$-term, all the gauge non-invariant terms will drop out from the low
energy effective theory. Only gauge invariant combinations can appear in the
effective theory.\cite{BFG0212,SM0204}

\begin{figure}
\centerline{
\includegraphics[width=2.8in]{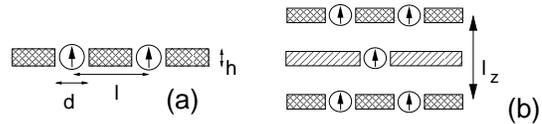}
}
\caption{
(a) A 2D device and (b) a 3D device made of electric dipoles screened by
superconducting films.
}
\label{sample}
\end{figure}

\section{Realistic devices}
\label{sec:real}

In the following, we will discuss how to design realistic devices that realize
our 2D and 3D models.  First we note that our 2D model Hamiltonian \Eq{model}
can be realized by magnetic or electric dipoles which form a Kagome lattice
(assuming only dipolar interactions between the dipoles).  For such a system
$t= S^2 U/2$, $t'=3S^2U/2$ and $J= -2U$.  So the coupling constants do not
have the right values to support an artificial light.  Thus the key to design
a working device is to find a way to reduce the couping between $S^{\pm}$.  We
need to reduce the $t$-term and $t'$-term by a factor $\sim 4S^2$.  We also
need to introduce an anisotropic spin term $(S^z)^2$ to bring  $J$ close to
zero.

We can use molecules with a finite electric dipole moment $\v d$ as our spins.
For a fixed $\v d$, the molecule should have two degenerate ground states with
angular momentum $\pm S$ in the $\v d$ direction. If we allow the molecule to
rotate, the ground states of the molecule will contain $2(2S+1)$ states
$|S^z,\si^z\>$, $S^z=-S,...,S$ and $\si^z=\pm 1$. $S^z$ corresponds to the
spin degree of freedom and $\si^z$ the isospin degree of freedom. The
tunneling between the $\pm S$ states generates a term $\del H=J'\si^x$ which
leads to quantum fluctuation of $\si^z$.
We also need to put the molecule, say, in a $C_{80}$ buckyball so the dipole can
rotate freely. We note that endohedral $S_cA_2N@C_{80}$\cite{SFH0027} is
commercially available from Luna Nanomaterials
(http://www.lunananomaterials.com), where $A$ is a rare earth atom such as $Y,
Er, Gd, ...$ If endohedral $S_cABN@C_{80}$ can be made with $A$ and $B$ being
different rare earth atoms, such an endohedral may have the properties
discussed above.

One way to reduce $t,t'$ is to embed dipoles in a fully gapped
superconductor.  A particular design for our 2D model is given in Fig.
\ref{sample}a.  The sample is  formed by a superconducting film.  Circular
holes of diameter $d$ are drilled through the film to form a Kagome lattice.
The dipoles are placed in the holes. A large $h$ will reduce $t$.  The
screening of the superconducting film also make the dipoles to tend to point
horizontally (\ie $S^z=0$).  In this case $J$ can be tuned by changing $d/l$.
If we choose $l=10$nm, $S=2$ and dipole moment $0.1e\cdot$nm, we find $U\sim
40$mK. The operating temperature to observe artificial light is about $1$mK,
which is achievable. 

The 3D model can be realized by the device in Fig. \ref{sample}b. We note that
the 3DK-lattice is formed by alternatively stacking K-lattices and triangular
lattices together. The top and the bottom layers in Fig. \ref{sample}b are
screened K-lattices just like Fig. \ref{sample}a, while the middle layer is a
screened triangular lattice. The distance between layers and $d/l$ need to be
tuned to reproduce the $U$-term. The $t$-term and $J$-term can be adjusted
similarly as in the 2D device.

\section{Physical properties of 2D and 3D devices}
\label{sec:phy}

The 2D and 3D devices are described by model Hamiltonian \Eq{model}
with coupling constants $U$, $J$, $t$ and $t'$. The low energy effective
theory \Eq{lgauge} contains only two  coupling constant $\t J$ and $g$ in
large $U$ limit.  $\t J$ and $g$ are determined by $U$, $J$, $t$ and $t'$.
If $U=40$mK, we can tune $t$ to make $g=6$mK. We can tune $J$ to make
$\t J = g/2=3$mK.

\begin{figure}
\centerline{
\includegraphics[width=2.8in]{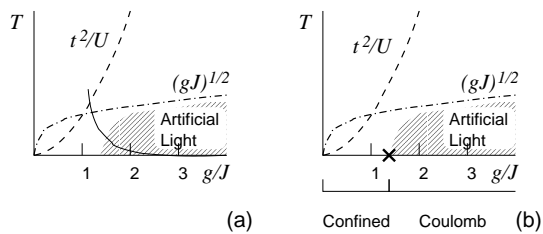}
}
\caption{
The phase diagram of (a) 2D and (b) 3D electric dipole systems.
The dash line represents the time reversal symmetry breaking transition.
The transition happens at
$T_c \sim t^2S^{-4}/U$. The dash-dot line marks the bandwidth of
artificial light, which is of order $\sqrt{\t J g}$. The thin solid line in
(a) marks the energy gap $\Del$ of the artificial light. The artificial light
exists in the shaded region where $T, \Del < \sqrt{\t J g}$.
In (b) the cross marks the position of a zero-temperature phase transition
between the confined phase and the Coulomb phase. 
The artificial light is exactly gapless in the Coulomb phase.
}
\label{phasedia}
\end{figure}

The phase diagrams of the 2D and 3D devices are sketched in Fig.
\ref{phasedia}. Both 2D and 3D electric dipole systems have a phase transition
at $T_c \sim t^2S^{-4}/U$ which breaks the time reversal symmetry.  The 3D
system also has a quantum phase transition at $g/\t J \sim 1$.  The quantum
phase transition separates the confined phase where the artificial light has
an energy gap and the Coulomb phase where the artificial light is gapless. In
principle, the quantum phase transition can be a continuous phase transition
although it does not change any symmetry.  We know that the Coulomb phase
corresponds to a phase with strong fluctuations of large closed strings. There
are two ways in which the Coulomb phase can change into the confined phase.
In the first way, the large closed string-nets break up into small open
string-nets.  This corresponds to condensation of charged bosons and produces
in a Anderson-Higgs phase. (Note that a confined phase is the same as a
Anderson-Higgs phase.) Such a transition is expected to be first
order.\cite{CW7388,HLM7492,P7822} In the second way, the large closed
string-nets start to cost too much energy and the ground state after
transition contains only dilute small closed string-nets.  Such a transition
was believed to be continuous.\cite{FS7982}

Both type of the transitions are between the same pair of phases - the Coulomb
and the confined phases. Both type of transition can appear in our 3D dipole
model. However, in the large $U$ limit, we expect the quantum phase transition
from the Coulomb phase to the confined phase to be the second type and to be
continuous. The continuous phase transition will become a smooth
cross over at finite temperatures (see Fig. \ref{phasedia}b). If $U$ is not
large enough, the quantum phase transition can be the first type which is a
first order phase transition. Such a first order phase transition will
extend to finite temperatures.

In the 2D electric dipole system, there is no zero-temperature quantum phase
transition and the artificial light always has a finite energy gap $\Del$ (see
\Eq{lightgap}).  The thin solid line in Fig. \ref{phasedia}a marks the scale
of the energy gap. When the energy gap is much less than the bandwidth
$\sqrt{\t J g}$ of the artificial light, we say the artificial light exists.

Our 2D and 3D dipole systems have boundaries. Some interesting questions
arise.  To the artificial light, what is the properties of the boundary? If we
shine artificial light onto the boundary, does artificial light get reflected
or absorbed? If we place an artificial charge near the boundary, whther the
charge is attracted or repelled by the boundary? Those questions can be
answered by our string-net picture of artificial light and artificial charges.
We note that the closed strings are always confined in the sample. The ends of
open strings always cost an energy of order $U$ even when the open string is
ended on the boundary. This means that the closed strings do not break up near
the boundary. Since the closed strings represent electric flux tube, we find
that the electric flux of the artificial light can never leave the sample
neither can end at the boundary of the sample. Therefore, to the artificial
light, the outside of the sample behaves like a perfect dia-electric media
which repels all the artificial electric flux.  If we place an artificial
charge near the boundary, the charge will be repelled by the boundary.

To understand the physical properties of the artificial light in the 2D model,
we can take the continuum limit by writing 
\begin{align}
 a_{\v i\v j} = & \del \v x_{\<\v i\v j\>} \cdot \v a(\v x) \nonumber\\
a_{0,\v i} = & a_0(\v x),
\end{align}
where $\v a = (a_x,a_y)$ is a 2D vector field (the vector gauge potential in
2D), $a_0$ corresponds to the potential field, $\v x$ is near the site $\v i$,
$\del \v x_{\<\v i\v j\>}$ is the vector that connect the $\v i$ and $\v j$
sites in the H-lattice, and $l$ is the distance between the neighboring sites
in the H-lattice.  In the continuum limit, the Lagrangian \Eq{lgauge} becomes
\begin{equation}
 L = \int d^2\v x \left( \frac{1}{4\t J\sqrt{3}} \v e^2 
   - \frac{3\sqrt{3}gl^2}{4} b^2
\right)
\end{equation}
where 
$\v e = \prt_t \v a - \prt_{\v x} a_0$ and
$\v b = \prt_x a_y - \prt_y a_x$
are the corresponding artificial electric field and artificial magnetic field.
We see that the velocity of our artificial light is $ c_a = \sqrt{ 9g\t
Jl^2/\hbar^2}  \sim \frac{\sqrt{t\t J}}{U} tl$.  If we take $\t J=3$mK,
$g=6$mK and $l=10$nm, we find the speed of the artificial light is about
$c_a=20$m/s. The band width of the artificial light is about $E_a=\pi c_a\hbar
/2l=20$mK. The gap of the artificial light is about $\Del\sim 0.03 E_a$.

From \Eq{Lcharged}, we find the continuum Lagrangian that describes the charged
particles in the 2D model (in the $U\gg t$ limit)
\begin{align}
 L=& \int d^2\v x \sum_{I=1,2} 
\left( \phi_I^\dag (i\prt_t - a_0 -U)\phi_I
- \frac{9tl^2}{2} |(\prt_i + i a_i)\phi_I|^2
\right.  \nonumber\\
 & +
\left. \bar \phi_I^\dag (i\prt_t + a_0 -U)\bar \phi_I
- \frac{9tl^2}{2} |(\prt_i - i a_i)\bar \phi_I|^2
\right)  
\end{align}
where $\phi_I$ describe the positively charged bosons, $\bar \phi_I$ describe
the negatively charged bosons, $\psi_1, \bar \psi_1$ describe the charged
boson on the even sites of the H-lattice, and $\psi_2, \bar \psi_2$ describe
the charged boson on the odd sites of the H-lattice.  It costs energy $2U$ to
create a pair of charged bosons.  The mass of the bosons is $m=(9tl^2)^{-1}$
and $mc_a^2=2.3$mK.  We would like to note that the boson velocity can be
larger than the speed of artificial light.  The potential energy between a
positive and a negative charge is $V(r)=\frac{\sqrt{3}\t J}{\pi}\ln r$.  A
bound state of a positive charge and a negative charge (an artificial atom)
has a size of order $l\sqrt{3\sqrt{3} \pi t/\t J} = 6.6 l$.  For each angular
momentum $m\hbar$, the lowest energy level of the artificial atom is of order
$\frac{\ln(m)\sqrt{3}\t J}{\pi}=1.7\ln(m)$mK.

For the 3D model, if the layer separation is $l_z$, we find the Lagrangian
in the continuum limit is given by
\begin{align}
 L =& \int d^3\v x  \frac{1}{4\t J\sqrt{3}l_z} 
(e_x^2+e_y^2+\frac{l_z^2}{l^2}e_z^2) \nonumber\\
& -\int d^3\v x 
\frac{3\sqrt{3}gl^2}{4l_z} (\frac{2l_z^2b_x^2}{3l^2}+\frac{2l_z^2b_y^2}{3l^2}
+b_z^2)
\end{align}
where $\v e$ and $\v b$ are the artificial electric field and 
artificial magnetic field in 3D. We see that, in general, the speed of
artificial light is different in different directions. For simplicity, we
choose $l_z=l$ and ignore the anisotropy in the speed of artificial light.
That is we work with the following simplified Lagrangian
\begin{align}
 L =& \int d^3\v x  \left( \frac{1}{4\t J\sqrt{3}l} \v e^2
-\frac{3\sqrt{3}gl}{4} \v b^2 \right)
\end{align}
The speed of artificial light is $c_a = \sqrt{ 9g\t Jl^2/\hbar^2}$.
If we take $\t J=3$mK,
$g=6$mK and $l=10$nm, we find 
$c_a=20$m/s.  The band width of the artificial
light is about $E_a=\pi c_a\hbar /2l=20$mK. 
The above 3D Lagrangian can be rewritten in a more standard form
\begin{align}
 L =& \int d^3\v x  \frac{1}{8\pi \al} 
\left( \frac{1}{c_a} \v e^2 - c_a \v b^2 \right)
\end{align}
where $\al=\frac{1}{2\pi}\sqrt{\t J/3g}=1/15$ is the artificial fine structure
constant.  The mass of the charged boson $m$ is of order $(9tl^2)^{-1}$ and
$mc_a^2\sim 2.3$mK.  The artificial atom has an energy level spacing
$\frac12mc_a^2\al^2\sim 0.01$mK and a size of order $1/\al mc_a\sim
\frac{6\pi\sqrt{3} t}{\t J} l= 87 l$.

In the following, we will discuss one experiment that can detect some of the
above properties in the 2D system. (Note it is easier to create a 2D device.)
If we place a tip of scanning tunneling microscope near an electric dipole, we
can induce the following coupling $ \del H = E(t) {S^+_{\v I} + h.c.}$ to the
electric dipole. $S^\pm_{\v I}$ flips a spin on a link which create a pair of
bosons on the two ends of the link. The two bosons
carry positive and negative artificial charges. If we measure the
high frequency capacitance of the tip,\cite{FGA0090} we can see peaks at the
energy levels of the artificial atom $\om = 2U+\frac{\ln(m)\sqrt{3}\t J}{\pi}=
(80+1.7\ln(m))$mK$=(1668+35\ln(m))$MHz. We also note that a AC voltage on the
tip at lower frequencies can generate artificial light.  However, the tip of
scanning tunneling microscope is not an efficient antenna to generate
artificial light.

From the above discussion, it is clear that the electric dipole systems, if
can be created, really provide a model for artificial light, artificial
charge, and artificial electromagnetic interaction in both two and three
dimensions.  We know that the $SU(N)$-spin model that realize 3D artificial
light, artificial electron and artificial proton\cite{Wlight} is not
realistic. The dipole systems discussed here contain only artificial light. It
would be very interesting to design a realistic device that has artificial
light, artificial electron and artificial proton. In that case, we can have an
artificial world sitting on our palm.

This research is supported by NSF Grant No. DMR--01--23156
and by NSF-MRSEC Grant No. DMR--02--13282.

\bibliography{/home/wen/bib/wencross,/home/wen/bib/all,/home/wen/bib/misc,/home/wen/bib/publst} 

\end{document}